\documentclass[pre]{revtex4}
\usepackage[cp1251]{inputenc}
\usepackage[english,russian]{babel}

\usepackage{graphicx}
\usepackage{dcolumn}
\usepackage{eucal}
\usepackage[dvips]{epsfig}
\usepackage{amssymb}
\usepackage{amsmath}

\begin{document}


\newcommand{\bs}{\boldsymbol}
\newcommand{\mbb}{\mathbb}
\newcommand{\mcal}{\mathcal}
\newcommand{\mfr}{\mathfrak}
\newcommand{\mrm}{\mathrm}

\newcommand{\ovl}{\overline}
\newcommand{\p}{\partial}

\renewcommand{\d}{\mrm{d}}
\newcommand{\lap}{\triangle}

\newcommand{\lan}{\bigl\langle}
\newcommand{\ran}{\bigl\rangle}

\newcommand{\bse}{\begin{subequations}}
\newcommand{\ese}{\end{subequations}}

\newcommand{\be}{\begin{eqnarray}}
\newcommand{\ee}{\end{eqnarray}}

\newcommand{\ga}{\alpha}
\newcommand{\gb}{\beta}
\newcommand{\gc}{\gamma}
\newcommand{\gd}{\delta}
\newcommand{\gr}{\rho}
\newcommand{\eps}{\epsilon}
\newcommand{\veps}{\varepsilon}
\newcommand{\gs}{\sigma}
\newcommand{\gf}{\varphi}
\newcommand{\go}{\omega}
\newcommand{\gl}{\lambda}

\renewcommand{\l}{\left}
\renewcommand{\r}{\right}


\title{\bf ON THE STATIC DIELECTRIC PERMITTIVITY FOR COULOMB SYSTEM IN THE LONG WAVELENGTH LIMIT
}
\author{V.B. Bobrov, S.A. Trigger}
\address{Joint\, Institute\, for\, High\, Temperatures, Russian\, Academy\,
of\, Sciences, 13/19, Izhorskaia Str., Moscow\, 125412, Russia;\\
email:\,satron@mail.ru}

\begin{abstract}
On the basis of the exact relations the general formula for the static dielectric permittivity $\varepsilon(q,0)$ for Coulomb system is found in the region of small wave vectors $q$. The obtained formula describes the dielectric function $\varepsilon(q,0)$ of the Coulomb system in both limits: in the "metallic" \,state and in the "dielectric" \, one. On this basis the determinations of the "apparent" \,dielectric and the "apparent"\, radius of screening are introduced. In the random phase approximation (RPA) the exact relations for the function  $\varepsilon(q,0)$ in the long-wavelength region of $q$ are found for an arbitrary degeneration of the particles.\\

PACS number(s): 52.25.Mq, 52.27.Gr, 52.25.Kn, 64.60.Bd\\

\end{abstract}

\maketitle

\section{Long wavelength dielectric function for Coulomb system: between dielectric and metallic states}

We consider the static dielectric permittivity $\varepsilon(q,0)$ for the homogeneous and isotropic Coulomb system. The function $\varepsilon(q,0)$ is determined as the proportionality coefficient between the potential $U^{tot}(q,0)$ of the total electric field in the medium and the external field potential $U^{ext}(q,0)$ [1]
\begin{eqnarray}
U^{tot}(q,0)=\frac{U^{ext}(q,0)}{\varepsilon(q,0)}. \label{A1}
\end{eqnarray}

According to [2,3] the function $\varepsilon(q,0)$ is connected with the static polarization operator $\Pi(q,0)$, which determines the response of the system on the screened external field by the relation
\begin{eqnarray}
\varepsilon(q,0)=1-\frac{4\pi}{q^2}\Pi(q,0), \label{A2}
\end{eqnarray}
\begin{eqnarray}
\Pi(q,0)=\sum_{a,b}z_a z_b e^2\Pi_{a,b}(q,0), \label{A3}
\end{eqnarray}
where $z_a e$, $m_a$ and $n_a$ are the charge, the mass and the average density of the particles of the sort $a$ in the system with the chemical potentials $\mu_a$ at temperature $T$. The system is considered under the condition of quasineutrality
\begin{eqnarray}
\sum_{a} e z_a n_a=0 \label{A4}
\end{eqnarray}

The functions $\Pi_{ab}(q,0)$ are the partial polarization operators of the particle species $a$ and $b$. In diagram technic [2,3] the functions $\Pi_{ab}(q,0)$  are the irreducible (on one-line of the Coulomb interaction in the q-channel) parts of the appropriate "density-density" Green functions $\chi_{ab}(q,0)$, which determine response of the system on an external field. In contrast with the Green functions $\chi_{ab}(q,0)$ of the Coulomb systems, the polarization functions $\Pi_{ab}(q,0)$ do not consist the singularities and are the smooth functions in the region of small wave vectors $q$ (at least for the normal systems). Therefore, the functions $\Pi_{ab}(q,0)$ for the small values of $q$ can be represented in the form
\begin{eqnarray}
\Pi_{a,b} (q,0)\simeq \pi_{a,b}^{(0)}+q^2 \pi_{a,b}^{(2)}, \label{A5}
\end{eqnarray}

\begin{eqnarray}
\pi_{a\,b}^{(0)}=lim_{q\rightarrow 0} \Pi_{a\,b} (q,0), \; \pi_{a\,b}^{(2)}=lim_{q\rightarrow 0} \left[\frac{\Pi_{a,b} (q,0)-lim_{q\rightarrow 0} \Pi_{a\,b} (q,0)}{q^2}\right]. \label{A6}
\end{eqnarray}

In [4,5] it was  shown that
\begin{eqnarray}
\pi_{a\,b}^{(0)}=-\left(\frac{\partial n_a}{\partial \mu_b}\right)_T. \label{A7}
\end{eqnarray}

Eq.~(\ref{A7}) is the generalization for the many-component Coulomb system the well known result [6] for the model of the one-component electron liquid, where this kind of equality names as "the sum rule for compressibility" [7]
\begin{eqnarray}
\pi_{e\,e}^{(0)}=-\left(\frac{\partial n_e}{\partial \mu_e}\right)_T=-n_e^2 K_T^e, \nonumber\\
K_T=-\frac{1}{V}\left(\frac{\partial V}{\partial P}\right)_T, \label{A8}
\end{eqnarray}
where V,  P and $K_T$ are respectively the volume, pressure and the isothermal compressibility of the Coulomb system.
In one's turn, for the two-component Coulomb system, which consists the electrons (index - $e$) and nuclei (index - $c$), the limiting relations [4,5] for the static structure factors $S_{a\,b} (q)$ can be found on the basis of  Eq.~(\ref{A8})
\begin{eqnarray}
lim \,S_{c\,c}(q\rightarrow 0)=n_c T K_T; \;\; lim \,S_{c\,c}(q\rightarrow 0)=\frac{n_c}{n_e}\,lim\, S_{e\,e}(q\rightarrow 0)=\left(\frac{n_c}{n_e}\right)^{1/2}lim \,S_{e\,c}(q\rightarrow 0) \label{A9}
\end{eqnarray}

The functions $S_{a\,b} (q)$ are measured directly, including the critical point region [8], in the experiments on the neutron scattering. By inserting (\ref{A5}), (\ref{A6}) in  (\ref{A2}) and (\ref{A3}) we obtain in the long-wavelength limit (the small values of $q$) for the dielectric function of the Coulomb system of an arbitrary composition the following relations
\begin{eqnarray}
\varepsilon(q,0)=\varepsilon_0^{st}+ \frac{\kappa^2}{q^2}, \;\;\; \kappa^2=-4\pi \sum_{a\,b}e^2 z_a z_b \pi_{a\,b}^{(0)}= 4\pi \sum_{a\,b}e^2 z_a z_b \left(\frac{\partial n_a}{\partial \mu_b}\right)_T  \label{A10}
\end{eqnarray}
\begin{eqnarray}
\varepsilon_0^{st}=1+4\pi \alpha, \; \;\; \alpha= -\sum_{a\,b}e^2 z_a z_b \pi_{a\,b}^{(2)}\label{A11}
\end{eqnarray}

It is evident, that all the coefficients in Eqs.~(\ref{A10}), (\ref{A11}) are the functions of the thermodynamic parameters of the Coulomb system. By use of the grand canonical ensemble one easily arrive [9] at the equality
\begin{eqnarray}
T \left(\frac{\partial n_a}{\partial \mu_b}\right)_T =\frac{1}{V} <\delta N_a \delta N_b>, \;\;\; \delta N_a =N_a - <N_a>, \label{A12}
\end{eqnarray}
where $n_a=<N_a>/V$, $N_a$ is the operator of the total number of particles of the sort $a$ and the brackets $<...>$ means the averaging on the grand canonical ensemble.

Inserting (\ref{A12}) in (\ref{A10}) and taking into account the quasineutrality condition we find for the Coulomb system at the arbitrary parameters
\begin{eqnarray}
\kappa^2=\frac{4\pi}{T}\,\frac{<Z^2>}{V}\geq 0, \; \;\; Z=\sum_{a} z_a e N_a. \label{A13}
\end{eqnarray}
We have mention that the sign of the value $\alpha$, which is introduced by Eq.~(\ref{A11}), is not determined in the moment. It is easy to see from (\ref{A11}) that the value $\kappa$ coincides in the appropriate limiting cases with the Debye and the Thomas-Fermi wave vectors. This value, as it follows from (\ref{A10}), characterizes the depth of penetration for the electromagnetic field in the medium. In the limiting case
\begin{eqnarray}
\frac{4\pi}{T}\,\frac{<Z^2>}{V}=0 \label{A14}
\end{eqnarray}
the penetration depth tends to infinity. Therefore, when the condition (\ref{A14})
is fulfilled, the Coulomb system manifests itself as an "apparent" dielectric, which changes only the amplitude of the electrostatic field on the value $\varepsilon_0^{st}$. In this sense the value $\varepsilon_0^{st}$ can be treated as the dielectric constant of the medium. Accordingly, the value $\alpha$ (\ref{A11}) has to be considered as the electric polarization of the medium.

It is necessary to stress the essential circumstance.
As in the case of "traditional" consideration of the "metal-dielectric" transition on basis of the analysis of
the electron conductivity (see, e.g., [10],[11]), one can maintain the relative character
of the matter division  on dielectrics and conductors. All dielectrics possess the non-zero conductivity for $T\neq 0$. The similar statement is applied to the depth of penetration in matter of the electrostatic field. This depth of penetration is determined by the value reciprocal to $<Z^2>/V$. In metals the penetration depth is very small, at the same time for dielectrics it can be of one order with the size of the system.

As an illustration, let us consider the action of the point charge on the infinite homogeneous Coulomb system. Then, taking into account (\ref{A1}), in $r$-space we obtain
\begin{eqnarray}
\frac{U^{tot}(r)}{U^{ext}(r)}=\frac{2}{\pi}\int_0^\infty \frac{d q}{q}\,\frac{sin(q\,r)}{\varepsilon (q,0)} \label{A15}
\end{eqnarray}
In the limit $r\rightarrow \infty$ from Eqs.~(\ref{A10}), (\ref{A15}) directly follows
\begin{eqnarray}
\frac{U^{tot}(r)}{U^{ext}(r)}=\frac{1}{\varepsilon_0^{st}}\exp(-r/R_{scr}),\;\;\; R_{scr}\equiv\left(\frac{\varepsilon _0^{st}}{\kappa^2}\right)^{1/2}, \label{A16}
\end{eqnarray}
where $R_{scr}$ is the electrostatic field penetration length in matter, or the screening radius, according to the terminology, accepted in the theory of non-ideal plasma (see, e.g., [12]). It is important to mention, that if the condition (\ref{A14}) is fulfilled for the "apparent" dielectric
\begin{eqnarray}
R_{scr}\rightarrow\infty. \label{A17}
\end{eqnarray}
Therefore, we can assert, that the representation of the dielectric permittivity $\varepsilon (q,0)$ in the region of small wave vectors $q$ in the form Eq.~(\ref{A10}) is universal and can be used for description of the Coulomb system in both, the "metallic"\, and the "dielectric"\, states of matter. The indirect confirmation of this statement contains in [3], where the generalized random phase approximation for determination of the polarization function is developed. This approximation permits to take into account the bound states of the electrons and nuclei.

\section{Dielectric constants of electron gas for arbitrary degeneration}

Let us show that for the consecutive consideration of the quantum effects the function $\varepsilon _0^{st}$ is not zero even for the ideal gas approximation for the polarization function $\Pi(q,0)$. Therefore the screening length $R_{scr}$ is different, in the appropriate approximations, from the Debye radius and from the Thomas-Fermi radius. Apparently, the first mention on this circumstance has been done in the papers [13],[14], devoted to calculation of the thermodynamic functions of the weakly non-ideal plasma. Following to the terminology introduced in [14] we name this value the "apparent" screening radius.

To illustrate the above statements we consider the static dielectric function $\varepsilon(q,0)$ in the random phase approximation for the electron gas on the positive charge compensation background. Then [3]
\begin{eqnarray}
\varepsilon^{RPA}(q,0)=1-\frac{4\pi}{q^2}\Pi^{RPA}(q,0), \label{A18}
\end{eqnarray}
\begin{eqnarray}
\Pi^{RPA}(q,0)=2\,\int \frac{d^3 p}{(2\pi)^3}
\frac{f_{{\bf p}-{\bf k}/2}-f_{{\bf p}+{\bf k}/2}}{\varepsilon_{{\bf p}-{\bf k}/2}-\varepsilon_{{\bf p}+{\bf k}/2}}=-\frac{m}{\pi^2\hbar^2 q}\int_0^\infty p f(p)\, ln\mid \frac{2p+q}{2p-q}\mid d p\leq 0 . \label{A19}
\end{eqnarray}
\begin{eqnarray}
\varepsilon(p)=\frac {\hbar^2 p^2}{2m},\;\;\; f(p)=\frac{1}{\exp ([\varepsilon(p)-\mu]/T) -1},\;\;\; n=2\,\int \frac{d^3 p}{(2\pi)^3} f(p) . \label{A20}
\end{eqnarray}
It is easy to check that from  Eqs.~(\ref{A19}), (\ref{A20}) directly follows Eq.~(\ref{A8}). In the two limiting cases: strong degeneration (DEG) and quasiclassical approximation (QCL), the integral (\ref{A19}) can be exactly calculated (see, e.g., [3]). In particular,
\begin{eqnarray}
\Pi^{DEG}(q,0)=-\,\frac{3 n}{2 \epsilon_F}\left\{\frac{1}{2}+\frac{k_F}{2 q}\left(1-\frac{q^2}{4k_F^2}ln\mid \frac{q+2k_F}{q-2k_F}\mid \right)\right\},\label{A21}
\end{eqnarray}
where $k_F$  and $\epsilon_F$ are the Fermi-vector and the Fermi-energy respectively. From Eq.~(\ref{A21}), taking into account Eqs.~(\ref{A10}), (\ref{A11}), one easy find
\begin{eqnarray}
\kappa^2=k_{TF}^2\,, \;\; \varepsilon_0^{st}=1-\frac{\gamma r_s}{3\pi}\,,\;\; k_{TF}=\left(\frac{6 \pi ne^2}{\epsilon_F}\right)^{1/2}, \;\; \gamma=\left(\frac{4}{9 \pi}\right)^{1/3}.\label{A22}
\end{eqnarray}

Here $k_{TF}$ is the Thomas-Fermi wave vector, $r_s$ is the known interaction parameter [7], which is determined as the ratio of the average distance between the particles to the Bohr radius $a_0=\hbar^2/2me^2$.
In this paper we don't discuss the problem of the Friedel oscillations (see, e.g., [7]).

For the quasiclassical case by use the Maxwellian distribution we arrive at the expression
\begin{eqnarray}
\Pi^{QCL}(q,0)=-\,\frac{n}{T} F_{1,1}\left(1,\, \frac{3}{2},\, - \frac{\hbar^2q^2}{8mT}\right);\label{A23}
\end{eqnarray}
\begin{eqnarray}
\kappa^2=k_D^2\,, \;\; \varepsilon_0^{st}=1-\frac{k_D^2 \lambda^2}{8}\,,\;\; k_D=\left(\frac{4 \pi ne^2}{T}\right)^{1/2}, \;\; \lambda=\left(\frac{\hbar^2}{mT}\right)^{1/2}.\label{A24}
\end{eqnarray}
Here $F_{1,1}\left(\alpha,\, \beta,\,z \right)$  is the degenerate hypergeometric function, $k_D$ is the Debye wave vector and $\lambda$  is the De Broigle wave length.

As it is shown above, the distinction $\varepsilon_0^{st}$ from unit leads to the difference between the value of the "apparent" screening radius Eq.~(\ref{A16}) and the usual expression for the screening radius in the theory of the weakly non-ideal Coulomb systems.
For numerical modeling of the various properties of Coulomb systems the relations between the different parameters
for an arbitrary generation of the electrons are often required. In the case under consideration these parameters are
$\kappa$ and $\varepsilon_0^{st}$ (or, according to (\ref{A5}), (\ref{A6}) the functions $\pi_{ee}^{(0)}$ and $\pi_{ee}^{(2)}$). This problem is actual, as an example, for investigations of the surface properties
of liquid metals [15]. For solution of this kind of problems it is necessary to make the expansion on the powers of the small wave vector $q$ under the integral in Eq.~(\ref{A19}). This expansion has the asymptotic character. In the framework of (\ref{A19}) the straightforward expansion leads to the divergent result for the region of small values of $p$. However, after integration by parts in (\ref{A19}) can be represented in the form
\begin{eqnarray}
\Pi^{RPA}(q,0)=\frac{1}{2\pi^2 q}\int_0^\infty \left\{\left(p^2-\frac{q^2}{4}\right) ln \mid \frac{2p+q}{2p-q}\mid +q p \right\} \frac{\partial f(p)}{\partial\varepsilon(p)}p \,d\, p. \label{A25}
\end{eqnarray}
In Eq.~(\ref{A25}) the expansion on small $q$ can be easily performed and we arrive at the necessary results
\begin{eqnarray}
\pi_{ee}^{(0)}=2\int \frac{\partial f(p)}{\partial\varepsilon(p)} \frac{d^3 p}{(2\pi)^3}=-2\int \left(\frac{\partial f(p)}{\partial \mu}\right)_T \frac{d^3 p}{(2\pi)^3}=-\left(\frac{\partial n}{\partial \mu}\right)_T \label{A26}
\end{eqnarray}
\begin{eqnarray}
\pi_{ee}^{(2)}=-\frac{1}{12\pi^2}\int \frac{\partial f(p)}{\partial\varepsilon(p)}\,d\, p=\frac{1}{12\pi^2}\left(\frac{\partial \varphi}{\partial \mu}\right)_T,\;\;\; \varphi(\mu,T)=\int_0^\infty f(p) d p. \label{A27}
\end{eqnarray}
Here the function $f(p)$ is determined by Eq.~(\ref{A20}). In the appropriate approximations for the distribution function the relations (\ref{A26}),(\ref{A27}) turn into the results (\ref{A22}),(\ref{A24}). The value $\alpha$, which is determined by Eq.~(\ref{A11}) is negative in the model under consideration for arbitrary degeneration of the electrons.

The authors are thankful to the Netherlands Organization for Scientific Research (NWO) for
support of this work in the framework of the grant № 047.017.2006.007.

\end{document}